\documentclass{aa}
\usepackage{graphicx}
\usepackage[mathscr]{eucal}
\begin{document}

\title{Wide field interferometric imaging with single-mode fibers}
       
\author{Olivier Guyon \inst{1,2}}

\offprints{Olivier Guyon}

\institute{Institute for Astronomy, University of Hawaii, 640 N. A'ohoku Place, Hilo, HI 96720 \and Universit\'e Pierre et Marie Curie, Paris 6, 4 Place Jussieu, 75005 Paris, France.}

\date{Received / Accepted }

\abstract{
Classical single-mode fiber interferometers, using one fiber per aperture, have very limited imaging capabilities and small field of view. Observations of extended sources (resolved by one aperture) cannot be fully corrected for wavefront aberrations: accurate measurements of object visibilities are then made very difficult from ground-based fiber interferometers. 
These limitations are very severe for the new generation of interferometers, which make use of large telescopes equipped with adaptive optics, but can be overcome by using several fibers per aperture. This technique improves the wide field imaging capabilities of both ground-based and space interferometers.
}

\maketitle

\section{Introduction}
In the visible and near-infrared, single-mode optical fibers allow efficient transport of coherent light over large distances: the transmission losses inside the fiber are typically 1dB/km in the near-infrared. The throughput of a fiber interferometer can be very high, since the number of optical elements before the fiber is small, and there is no need for numerous beam steering mirrors. Moreover, the beam combination can be done using fiber optics couplers, and does not require high numbers of glass to air interfaces.
A second interesting property of single mode optical fibers is the spatial filtering of the incoming wavefronts. Spatial filtering with single-mode optical fibers is superior to pinhole filtering (\cite{kee01}) and is used in interferometry to restore coherence of the telescope's beams. Single-mode fibers only allow the propagation of a fundamental mode ($LP_{01}$). The loss of light coherence of the entrance pupil due to wavefront aberrations is then traded for a loss of coupling efficiency (\cite{vcf00}). The object visibility measurements can be corrected for this loss of coupling efficiency by recording in real time the amount of light coupled in each fiber (real-time self calibration of the interferometer), which greatly improves the accuracy of the object visibility measurement.
Thanks to these two properties, fiber optics offer a reliable and simple solution to transport and combine coherent beams in interferometers with few apertures, and the recent progress in integrated optics will soon allow complex beam combiners to be built for larger interferometer arrays (\cite{mal99,ber99,ber01}).

Unfortunately, the restoration of coherence by spatial filtering comes at the expense of a very small field of view (FOV), which is the size of the diffraction spot of the largest telescope in the array. This is a serious limitation for the new generation of interferometers (Keck, VLTI, OHANA) which make use of large AO-equipped telescopes: the FOV is very small for such large telescopes. Moreover, these interferometers make it possible to observe fainter non-stellar sources such as AGNs and YSOs, for which the required FOV is often larger than for stellar observations. In this work, it is shown in \S2 and \S3 that spatial filtering reduces the field of view and that wavefront aberrations corrupt the object visibility measurements. This later effect is quantified and discussed in \S4 and \S5. The use of multiple fibers per telescope offers an attractive solution to these limitations (\S6) and greatly improves the wide field imaging capabilities of space (\S7) and ground-based (\S8) interferometers.

\section{Coupling efficiency and field of view}

\subsection{Coupling efficiency}
Coupling of starlight into a single-mode fiber is achieved by positioning the fiber head in the focal plane and centering it on the PSF of the star. The coupling efficiency, which is the fraction of the starlight coupled into the fiber, is given by
\begin{equation}
\rho = \frac{\left|\int E_{fp} E^*_{01} dA\right|^2}{\int \left| E_{fp} \right|^2 dA \times \int \left| E_{01} \right|^2 dA},
\end{equation}
where $E_{fp}$ is the incident electric field at the head of the fiber, in the focal plane, and $E_{01}$ is the electric field of the fundamental ($LP_{01}$) mode of propagation in the fiber. $E_{fp}$ is obtained by Fourier transform of the electric field in the entrance pupil, and $E_{01}$ is very well approximated by a Gaussian of constant phase across the width of the fiber. The coupling efficiency between the $LP_{01}$ mode and the Airy function is 0.82 (\cite{sha88,rui98}). Therefore, under optimal conditions, no more than 82\% of the incoming starlight can be coupled into the single-mode fiber. Fresnel reflection by the surface of the fiber head reduces this coupling efficiency to 78 \% unless an anti-reflective coating is used on the fiber head.\\

\subsection{Field of view of the fiber}
The coupling efficiency decreases rapidly with the distance of the point source to the optical axis of the telescope. Using equation (1), Shaklan and Roddier (1988) have shown that the coupling efficiency, in the case of a circular aperture and a fundamental mode approximated by a Gaussian, is
\begin{eqnarray}
\lefteqn{\rho(\vec{\alpha},d,f,\omega,\lambda) = \frac{8}{\omega^2} e^{-2\left(\frac{|\vec{\alpha}| f}{\omega}\right)^2}}\nonumber \\
 & \times \left[ \int e^{-\left(\frac{r}{\omega}\right)^2} I_0(2r|\vec{\alpha}| f/\omega^2) J_1(\pi dr/\lambda f) dr \right]^2.
\end{eqnarray}
Where $f$ is the telescope focal length, $d$ is the telescope diameter, $\lambda$ is the wavelength, $\omega$ is the $1/e$ width of the fundamental mode, $\vec{\alpha}$ is the angular position of the point source relative to the optical axis, $J_1$ is the first order Bessel function and $I_0$ is the zero order modified Bessel function. We note that this expression is only a function of $X_1 = |\vec{\alpha}|/(\lambda/d)$ and $X_2 = (f\lambda)/(\omega d)$ :

\begin{eqnarray}
\lefteqn{\rho(X_1,X_2) = 8\:e^{-2(X_1 X_2)^2}}\nonumber \\
 & \times \left[ \int e^{-r^2} I_0(2X_1 X_2 r) J_1(\frac{\pi r}{X_2}) dr \right]^2
\end{eqnarray}
where $r$ is now dimensionless. $\rho(0,X_2)$ is maximum for $X_2 = 1.401$, and this value will be used in this work. We can then express $\rho$ as a function of $X_1$ only, which leads to :
\begin{eqnarray}
\lefteqn{\rho(\vec{\alpha}) = 8\:e^{-3.923 \: \left(\frac{|\vec{\alpha}|}{\lambda/d}\right)^2} }\nonumber \\
 & \times \left[ \int e^{-r^2} I_0(2.802 \:\frac{|\vec{\alpha}|}{\lambda/d} r) J_1(\frac{\pi r}{1.402}) dr \right]^2.
\end{eqnarray}

The relation between the amplitude $E$ of the electric field coupled in the fiber and the luminosity $\mathcal{I}$ of a point source which position is $\vec{\alpha}$ is then
\begin{equation}
E(\vec{\alpha}) = \sqrt{\mathcal{I}} \frac{d \sqrt{\pi}}{2} \sqrt{\rho(\vec{\alpha})}.
\end{equation}
Figure 1 shows the coupling efficiency as a function of $|\vec{\alpha}|$, in units of $\lambda/d$, for an unobstructed circular pupil and a circular pupil with an central obscuration of diameter $0.4 \times d$. In each case, $f$ has been chosen to maximize the coupling efficiency for $|\vec{\alpha}| = 0$. At $|\vec{\alpha}|=\lambda/d$, the coupling efficiency is 10 times smaller than it is on-axis for a non-obstructed circular aperture, and $\rho(0.58 \lambda/d) = 0.5 \rho(0)$.\\

\begin{figure}
\centering
\includegraphics[width=9cm]{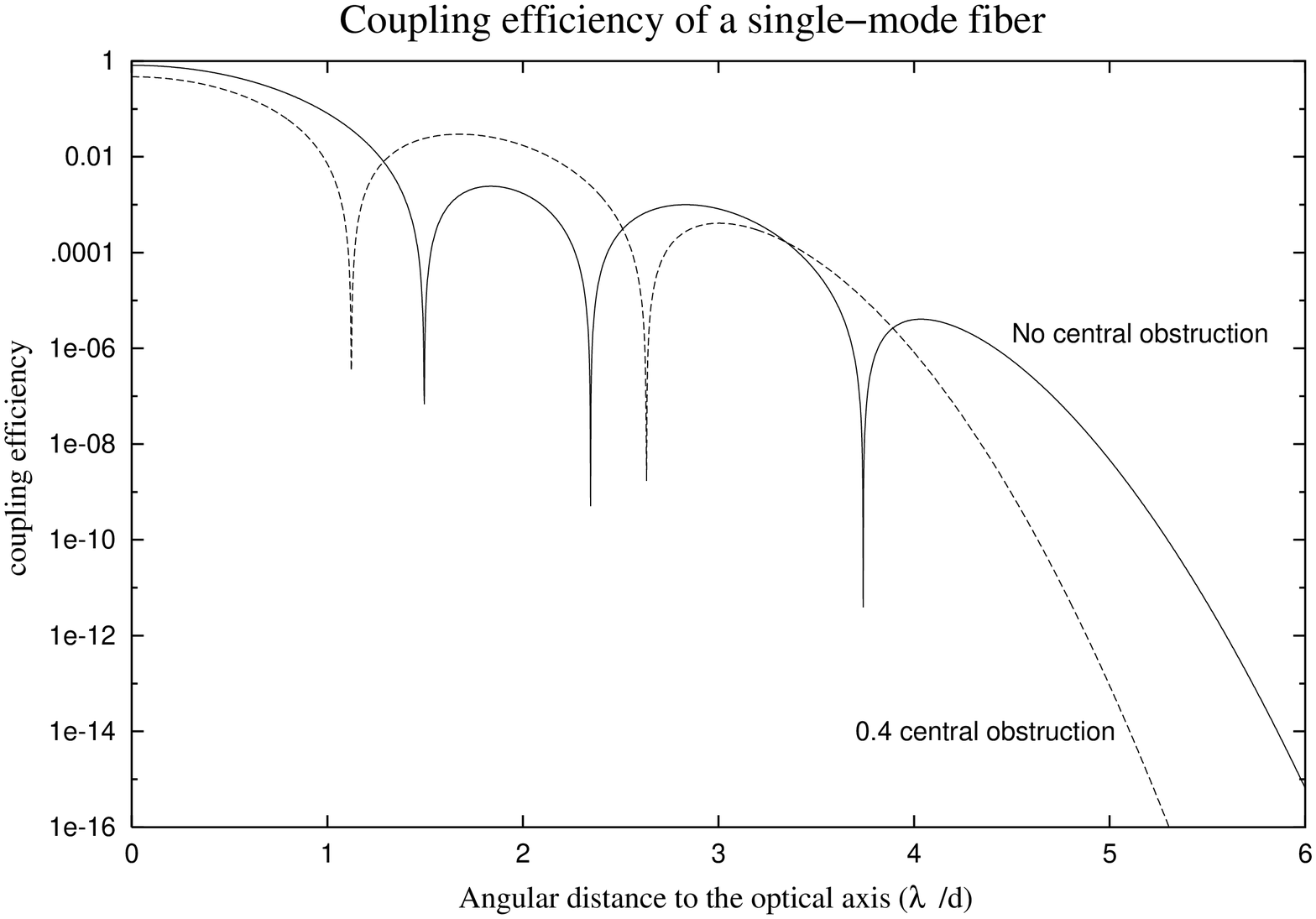}
\caption{The coupling efficiency of the light from a point source to a single-mode fiber at the telescope focus is a function of the position of the source on the sky. }
\end{figure}

Therefore, the field of view of an interferometer using one single-mode fiber per aperture to transport the beam is limited to $\lambda/d$. This very limited field of view (FOV) is a serious constraint for the observation of many objects, such as active galactic nuclei (AGNs) and young stellar objects (YSOs). Aperture synthesis using either the Earth rotation for ground-based interferometers or a rotation of the array for space interferometers could allow reconstruction of images with a much larger FOV than permitted by the use of single-mode fibers.

A point of concern is the use of single-mode fibers on an interferometer whose telescopes have different diameters and/or pupil shapes. The coupling efficiency, as projected on the sky, is then different for each telescope, and this effect has to be taken into account for image reconstruction of objects larger than $\lambda/d_{max}$ ($d_{max}$ is the diameter of the largest aperture in the array). Residual wavefront aberrations also produce a similar discrepancy between the fiber coupling efficiencies. These effects are studied in the next 2 sections.

\section{Observation of extended sources with a fiber interferometer}

\subsection{Instrumental response to a point source}
A general description of fiber interferometers is given in Coude du Foresto et al. (1997), and we only recall here some of the basic concepts for a 2 telescope interferometer (Fig. 2).

A fiber interferometer cophased (zero optical pathlength difference (OPD) from the source to the detector) for a direction $\vec{r_0}$ on the sky couples the electric field of the observed source on the telescopes' focal planes into the fibers. With a 2 telescope interferometer observing an unresolved point source at a direction $\vec{r}$, the electric fields $E_1$ and $E_2$ in the fibers, as a function of time $t$, are given by
\begin{equation}
\mathbf{E_i}(t) = E_i\:e^{j(\vec{\alpha} \cdot \vec{k_i}+\omega t)} 
\end{equation}
for $i=1,2$, with $\omega = 2\pi c/\lambda$, $\vec{\alpha} = \vec{r}-\vec{r_0}$ and $\vec{k_1}$,$\vec{k_2}$ are the positions of the 2 telescopes. The beam combiner (a fiber coupler in this case, noted X in Fig. 2) sums the two electric fields, with a time-dependent phase term $\psi(t)$ between the two, and the electric field at its output is 
\begin{eqnarray}
\mathbf{E}(t) & = & \mathbf{E_1}(t) + \mathbf{E_2}(t)e^{j\psi(t)} \nonumber\\
 & = & E_1\:e^{j(\vec{\alpha} \cdot \vec{k_1}+\omega t)} + E_2\:e^{j(\vec{\alpha} \cdot \vec{k_2}+\omega t+\psi(t))}.
\end{eqnarray}
The intensity measured at the output of the beam combiner is
\begin{equation}
I(t) = \mathbf{E}(t)\mathbf{E}^*(t)
\end{equation}
\begin{equation}
I(t) = E_1^2 + E_2^2 + 2 E_1 E_2\: cos(\vec{\alpha} \cdot \vec{k}+\psi(t)),
\end{equation}

\begin{figure}
\centering
   \includegraphics[width=8cm]{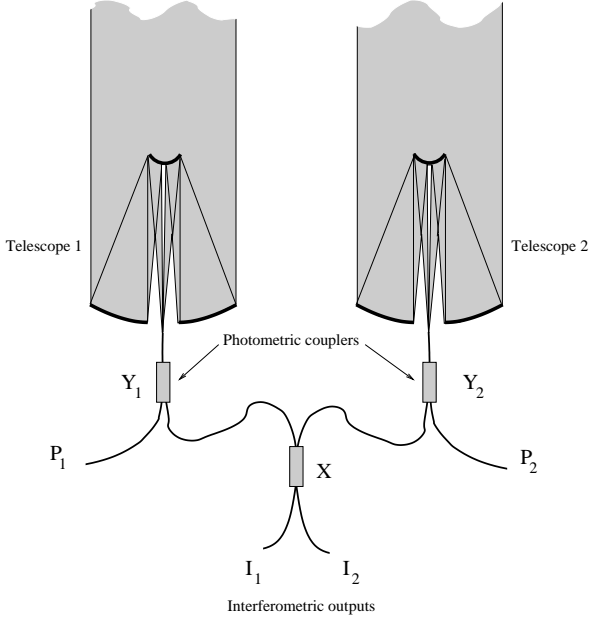}
   \caption{Optical layout of a 2 telescopes fiber interferometer. Adapted from \cite{cdf97}. The light of each telescope is coupled into a fiber in the focal plane. Before interferometric combination in fiber coupler $X$, the coupling efficiency in each arm of the interferometer is measured (photometric outputs $P_1$ and $P_2$) thanks to the fiber couplers $Y_1$ and $Y_2$.}
   \label{fiber_layout} 
\end{figure}

with the baseline vector $\vec{k} = \vec{k_1}-\vec{k_2}$. If a 50/50 fiber splitter is used, there are in fact two interferometric outputs (Fig. 2) :
\begin{equation}
I_1(t) = \frac{E_1^2 + E_2^2}{2} + E_1 E_2\: cos(\vec{\alpha} \cdot \vec{k}+\psi(t))
\end{equation}
and
\begin{equation}
I_2(t) = \frac{E_1^2 + E_2^2}{2} - E_1 E_2\: cos(\vec{\alpha} \cdot \vec{k}+\psi(t)).
\end{equation}

The OPD between the two beams is usually modulated ($\psi(t)$ is not constant), and the temporal modulation of $I$ obtained by this changing OPD leads to the measurement of $E_1^2 + E_2^2$ (mean level), $E_1 E_2$ (fringe amplitude) and $\phi = \vec{\alpha} \cdot \vec{k}$ (fringe phase).For example, most interferometers introduce a time-dependent OPD variation to scan the fringe packet and extract the fringe parameters in each scan.

The phase of the interferometric signal, $\phi = \vec{\alpha} \cdot \vec{k}$, can only be referenced through the simultaneous observation of the object and a reference star or by using phase closure techniques, which requires at least 3 apertures.

\subsection{Instrumental response to an extended source}
The expression of the interferometric signal for an extended source, of light distribution $\mathcal{I}(\vec{\alpha})$, is obtained by integration of equation (9) over $\vec{\alpha}$ :
\begin{equation}
I(t) = \int_{\vec{\alpha}} E_1^2(\vec{\alpha}) + E_2^2(\vec{\alpha}) + 2 E_1(\vec{\alpha}) E_2(\vec{\alpha})\: cos(\vec{\alpha} \cdot \vec{k}+\psi(t)) d\vec{\alpha}.
\end{equation}
From equation (5),
\begin{equation}
E_i(\vec{\alpha}) = \sqrt{\mathcal{I}(\vec{\alpha})} \frac{d_i \sqrt{\pi}}{2} \sqrt{\rho_i(\vec{\alpha})}
\end{equation}

for $i=1,2$, where $\rho_1$ and $\rho_2$ are the coupling efficiencies functions for telescopes 1 and 2 respectively (given by equation (4) for non-obstructed circular pupils without wavefront aberrations).
Therefore,
\begin{eqnarray}
\lefteqn{I(t) = \int_{\vec{\alpha}} \mathcal{I}(\vec{\alpha})\bigg[ \frac{d_1^2 \pi}{4} \rho_1(\vec{\alpha}) + \frac{d_2^2 \pi}{4} \rho_2(\vec{\alpha})}\nonumber \\
 & + 2 \frac{d_1 d_2 \pi}{4}\sqrt{\rho_1(\vec{\alpha}) \rho_2(\vec{\alpha})} \: cos(\vec{\alpha} \cdot \vec{k}+\psi(t)) d\alpha \bigg] 
\end{eqnarray}

\subsection{Measurements of the object's Fourier transform without wavefront aberrations}
Several cases are briefly discussed in the absence of wavefront aberrations :
\begin{enumerate}
\item {$d_1 = d_2 = d$, $\rho(\vec{\alpha})=\rho_0$}\\
The coupling of the fiber is considered constant across the field. This corresponds to the small field approximation of fiber interferometers, a valid approximation when the object is unresolved by the individual apertures. Equation (14) becomes
\begin{equation}
I(t) = \rho_0 \frac{d^2 \pi}{2}\left[ \int \mathcal{I}(\vec{\alpha}) d \vec{\alpha} + \mathcal{R}e \left[ e^{j\psi(t)} \tilde{\mathcal{I}}(\vec{k}) \right] \right]
\end{equation}
where $\tilde{\mathcal{I}}$ is the Fourier transform of $\mathcal{I}$. The interferometer measures one component of the Fourier transform of the source's light distribution per baseline. 

\item{$d_1 = d_2 = d$}\\
Because the two apertures have the same diameter, the coupling efficiency is also the same for the two apertures for all values of $\vec{\alpha}$. Equation (14) becomes
\begin{equation}
I(t) = \frac{d^2 \pi}{2}\left[ \int \mathcal{I}_{a}(\vec{\alpha}) d \vec{\alpha} + \mathcal{R}e \left[ e^{j\psi(t)} \tilde{\mathcal{I}}_{a}(\vec{k}) \right] \right]
\end{equation}
where
\begin{equation}
\mathcal{I}_{a}(\vec{\alpha}) = \rho(\vec{\alpha}) \times \mathcal{I}(\vec{\alpha}).
\end{equation}
The interferometer measures here one Fourier component (per baseline) of an apodized image of the source. The apodization profile is the transmission profile of the fiber on the sky.\\

\item{$d_1 \neq d_2$}\\
The coupling efficiencies of the two fibers on the sky are different. Equation (14) becomes
\begin{eqnarray}
\lefteqn{I(t) = \frac{d_1^2 \pi}{4}\left[ \int \mathcal{I}_{a1}(\vec{\alpha}) d \vec{\alpha} \right]}\nonumber \\
 & + \frac{d_2^2 \pi}{4}\left[ \int \mathcal{I}_{a2}(\vec{\alpha}) d \vec{\alpha} \right] + \frac{d_1 d_2 \pi}{2}\mathcal{R}e \left[ e^{j\psi(t)} \tilde{\mathcal{I}}_{a12}(\vec{k}) \right] \
\end{eqnarray}
where
\begin{equation}
\mathcal{I}_{ai}(\vec{\alpha}) = \rho_i(\vec{\alpha}) \times \mathcal{I}(\vec{\alpha}),
\end{equation}
for $i=1,2$, and
\begin{equation}
\mathcal{I}_{a12}(\vec{\alpha}) = \sqrt{\rho_1(\vec{\alpha}) \rho_2(\vec{\alpha})} \times \mathcal{I}(\vec{\alpha}).
\end{equation}
The interferometer measures a Fourier component of the product of the source and an apodization mask. The apodization mask is the square root of the product of the two fiber apodization profiles on the sky.
\end{enumerate}

\subsection{Effect of time-variable wavefront perturbations}
Time-variable wavefront aberrations change the fiber coupling efficiencies of each fiber for an on-axis point source. More seriously, the on-sky maps of the fiber coupling efficiency are changing in structure. In the following discussion, we note $\rho_1(\vec{\alpha},t)$ and $\rho_2(\vec{\alpha},t)$ the coupling efficiencies of the two fibers for the sky position $\vec{\alpha}$ at time $t$.\\

\begin{figure}
\centering
\includegraphics[width=8cm]{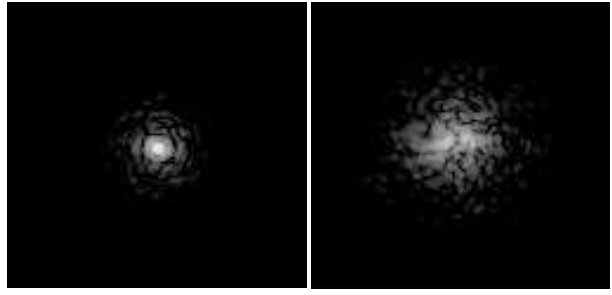}
\caption{Effect of the atmospheric turbulence on the coupling efficiency of a single-mode fiber on the sky. This figure shows typical instantaneous coupling efficiencies caused by atmospheric turbulence for two values of $d/r_0$ ($d$ is the telescope diameter and $r_0$ is the Fried parameter defining the strength of atmospheric turbulence). The brightness scale is logarithmic from 0 (no transmission) to 1 (total transmission).}
\end{figure}

\subsubsection{Observation of an object of small size}
Here, we consider an object which is not resolved by the individual apertures : the size of the object is much smaller than the diffraction limit of each telescope. The coupling efficiency is then constant across the object for each aperture. Because $\rho_1$ and $\rho_2$ are not functions of $\vec{\alpha}$, equation (14) becomes :
\begin{eqnarray}
\lefteqn{I(t) = \frac{d_1^2 \pi}{4}\rho_1(t) \int \mathcal{I}(\vec{\alpha}) d \vec{\alpha} + \frac{d_2^2 \pi}{4}\rho_2(t) \int \mathcal{I}(\vec{\alpha}) d \vec{\alpha}}\nonumber \\
 &  + \frac{d_1 d_2 \pi}{2} \sqrt{\rho_1(t) \rho_2(t)} \mathcal{R}e \left[ e^{j\psi(t)} \tilde{\mathcal{I}}(\vec{k}) \right].
\end{eqnarray}
The complex visibility of the observed object, $V(\vec{k})$, can be expressed as
\begin{equation}
V(\vec{k}) = \frac{\tilde{\mathcal{I}}(\vec{k})}{\tilde{\mathcal{I}}(\vec{0})}.
\end{equation}
The photometric outputs (Fig. 2), $P_1(t)$ and $P_2(t)$, are measuring respectively $\rho_1(t)$ and $\rho_2(t)$:
\begin{equation}
P_i(t) = \beta_i \frac{d_i^2 \pi}{4}\rho_i(t) \int \mathcal{I}(\vec{\alpha}) d \vec{\alpha}, 
\end{equation}
for $i=1,2$, where $\beta_1$ and  $\beta_2$ are the fraction of fiber flux being sent into the photometric outputs by the fiber splitter $Y_1$ and $Y_2$ respectively. $Q_1$ and $Q_2$ represent the flux in each fiber to the combiner (fiber splitter X).
\begin{equation}
Q_i(t) = \frac{1-\beta_i}{\beta_i} \times P_i(t)
\end{equation}
for $i=1,2$.
\begin{equation}
I(t) = Q_1(t) + Q_2(t) + \sqrt{Q_1(t) \times Q_2(t)} \times \mathcal{R}e \left[ e^{j\psi(t)} V(\vec{k}) \right].
\end{equation}
From equations (24) and (25), it appears that $|V(\vec{k})|$ can be computed from the photometric and interferometric output(s) $P_1(t)$, $P_2(t)$ and $I(t)$. Precise measurement of $\psi(t)$ is required to compute the phase of $V(\vec{k})$. This self-referencing technique is used on the fiber-fed FLUOR beam combination unit of the IOTA interferometer.

\subsubsection{Observation of an extended object}
If the size of the object is a significant fraction of the diffraction limit of the telescopes, $\rho$ is not uniform across the object: for a given telescope, the fiber coupling efficiency is a function of both the angular vector $\vec{\alpha}$ and of time. Figure 3 shows a typical instantaneous coupling efficiency $\rho$ as a function of position on the sky $\vec{\alpha}$ for two values of atmospheric turbulence.
The photometric outputs are:
\begin{equation}
P_i(t) = \beta_i \frac{d_i^2 \pi}{4}\int \rho_i(\vec{\alpha},t) \mathcal{I}(\vec{\alpha}) d \vec{\alpha}, 
\end{equation}
for $i=1,2$.
The interferometric output is:
\begin{equation}
I(t) = Q_1(t) + Q_2(t) + \sqrt{Q_1(t) Q_2(t)} \mathcal{R}e \left[ e^{j\psi(t)} \frac{\tilde{\mathcal{I}}_{a12}(\vec{k})}{\tilde{\mathcal{I}}_{a12}(\vec{0})} \right]
\end{equation}
where
\begin{equation}
\mathcal{I}_{a12}(\vec{\alpha}) = I(\vec{\alpha}) \sqrt{\rho_1(\vec{\alpha},t) \rho_2(\vec{\alpha},t)}.
\end{equation}
In this case, the knowledge of $I(t)$, $P_1(t)$ and $P_2(t)$ is not sufficient to compute the visibility $|V(\vec{k})|$ of the object: $\rho_1(\vec{\alpha},t)$ and $\rho_2(\vec{\alpha},t)$ also need to be known to accurately interpret the observed fringe visibility. To illustrate this effect, the simple example of the observation of a double star observed with two identical telescopes is considered. The wavefront of each telescope is affected by a random tip-tilt, due to atmospheric turbulence or telescope pointing errors. $\vec{\alpha_1}$ and $\vec{\alpha_2}$ are the positions of the two stars, of identical brightness and both unresolved by the interferometer. Three cases for an observation at time $t_0$ are considered:\\
{\bf (1)} Telescope 1 is pointing at star 1 and telescope 2 is pointing at star 2 ($\rho_1(\vec{\alpha_1},t_0) =1$, $\rho_1(\vec{\alpha_2},t_0) =0$, $\rho_2(\vec{\alpha_1},t_0) =0$ and $\rho_2(\vec{\alpha_2},t_0) =1$).\\
{\bf (2)} Telescope 1 and 2 are both pointing halfway between star 1 and 2 ($\rho_1(\vec{\alpha_1},t_0) = \rho_1(\vec{\alpha_2},t_0) = \rho_2(\vec{\alpha_1},t_0) = \rho_2(\vec{\alpha_2},t_0) = 0.5$).\\
{\bf (3)} Telescope 1 and 2 are both pointing at star 1 (($\rho_1(\vec{\alpha_1},t_0) = \rho_2(\vec{\alpha_1},t_0) = 1$, $\rho_1(\vec{\alpha_2},t_0) = \rho_2(\vec{\alpha_2},t_0) = 0$).\\
$P_1(t_0)$ and $P_2(t_0)$ are identical in each case : it is impossible to tell between the 3 cases from the values of the photometric signals. In case 1 there is no coherence between the light in the two fibers and $\mathcal{I}_{a12} = 0$ and the measured fringe visibility is 0, while in case 2, $\mathcal{I}_{a12} = 0.5 \times \mathcal{I}$ and the measured fringe visibility corresponds to the separation of the 2 stars. Finally, in case 3, the observed visibility is 1, because the interferometer behaves just as if there was only one star. During real observations, the atmospheric seeing and pointing errors of the telescopes would induce variations into the measured visibilities which cannot be corrected by measurements of $P_1$ and $P_2$.

\section{Effect of atmospheric turbulence on object visibility measurements.}

In this section, the amplitude of the effect presented above is estimated for the observation of various astrophysical sources. In these numerical simulations, only the effect of atmospheric turbulence has been simulated, for various values of $r_0$, the Fried parameter. This is adequate for ground-based interferometers without Adaptive Optics (such as CHARA, PTI, IOTA and NPOI). The effect of atmospheric turbulence on ground-based interferometers with adaptive optics (such as Keck, OHANA and VLTI) can still be obtained from this study by adopting a $r_0$ value which best represents the corrected wavefront: the value of $r_0$ that would yield the full width at half maximum (FWHM) of the corrected long-exposure PSF if there was no AO system. For each telescope+AO system, this derived value of $r_0$ will decrease as a function of the brightness of the source used for wavefront sensing.

To reproduce the effect of atmospheric turbulence, a simple 2-telescope fiber-fed interferometer has been simulated. Two independent time series of 2000 wavefronts ($d/r_0=1$) were first generated, one for each telescope. These wavefronts are then scaled to the appropriate value of $d/r_0$ for the simulation.
For each step, the simulator, written in C, first computes the on-sky fiber coupling efficiency map for each telescope. These two maps are then multiplied by the intensity map of the source to yield the values of the photometric outputs (equation 26). The square root of the product of these two maps is multiplied by the intensity map of the source (equation 28), and the result is then Fourier transformed to obtain the uncorrected fringe visibility, which is then corrected using the values of the photometric outputs (equation 27). Although the code can work with different telescope pupils (shape, diameter, central obstruction etc...), the results presented in this work were obtained with two circular pupils (no central obstruction) of identical diameter.

\begin{figure}
\includegraphics[width=8cm]{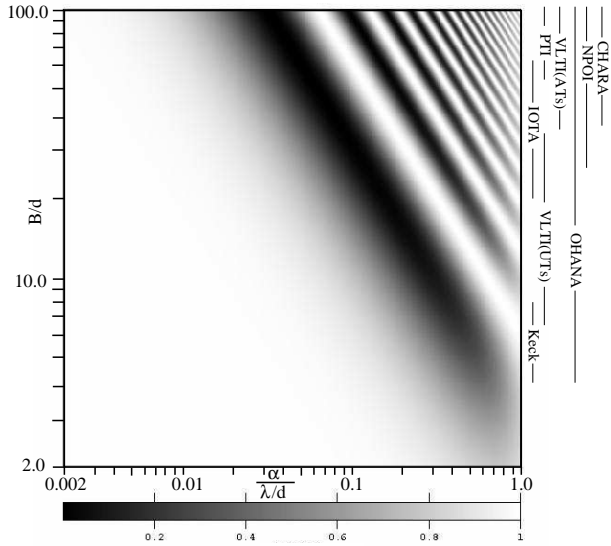}
\caption{Fringe visibilities for the observation of two stars of equal brightness separated by $\alpha$ (angular separation) without wavefront degradation ($d/r_0=0$). $B$ is the distance between the 2 telescopes of diameter $d$. $\lambda$ is the wavelength of the observation. On the right, $B/d$ values for some interferometers have been indicated.}
\end{figure}

\subsection{One example : the observation of a double star}
Mass estimates of stars rely upon precise measurements of orbits of double stars. Thanks to interferometers, it is possible to make such measurements on otherwise unresolved binaries (spectroscopic binaries).

In this simulation, one of the stars is on the optical axis of each telescope. As can be seen on Fig. 4, when the second star is close to the edge of the field of view of the fiber ($\alpha/(\lambda/d)$ is close to 1), the measured visibility is affected by the vignetting introduced by the fiber. One can correct for this effect because the electric field $E_{01}$ of the fundamental mode of the fiber is known. Atmospheric turbulence will introduce a variation (as a function of time) of the measured visibilities. One might think that it is possible to correct for this effect by averaging a large number of measurements. Figure 5 shows that such an average (here, an average of 2000 visibility measurements, each one of those has been previously been corrected according to the values of the photometric outputs $P_1$ and $P_2$) is significantly different from the measurement that would be obtained without atmospheric turbulence. The difference is especially large when the visibility of the object is low (the interferometer is resolving the double star): the error can then be larger than the measured value of the visibility. Figure 6 shows how these errors affect the measure of the system separation.
\begin{figure}
\includegraphics[width=8cm]{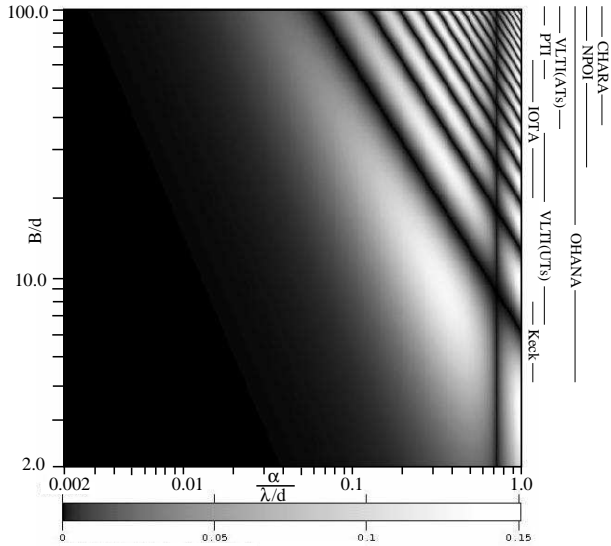}
\caption{This figure shows the absolute value of the mean difference between the fringe visibilities measurements with $d/r_0=1$ and the fringe visibilities measurements with $d/r_0=0$ (Fig. 4).}
\end{figure}

For $d/r_0$ values larger than 1, the errors are almost independent of $d/r_0$. This is an expected result since the error is then a function of the ratio between the size of the object (separation between the 2 stars in this example) and the size of a speckle: this ratio defines the gradient of fiber coupling efficiency across the object, which is the cause of this effect. Figure 7 shows how those errors vary as a function of $d/r_0$ for $\alpha/(\lambda/d) = 0.22$ and $B/d = 5.92$. Errors for other values of $\alpha/(\lambda/d)$ and $B/d$ vary in a similar way. It is found that the standard deviation between measurements is very close to the systematic error introduced by the atmospheric turbulence. This is true for all points except the ones close to $\alpha/(\lambda/d) = 0.7$, for which, although the standard deviation is relatively high, the average visibility measurement is not significantly affected by atmospheric turbulence. This ``equilibrium'' value of $\alpha/(\lambda/d)$ is a function of $d/r_0$ and is therefore not constant during an observation: this property cannot be used to hope to obtain exact visibility measurements. It is found that, in order to significantly decrease the visibility measurement error, $d/r_0$ needs to be smaller than 1. Existing Adaptive Optics systems deliver Strehl ratios of typically 30\% to 70\% on bright sources in the near-infrared. This degree of correction, as suggested in Fig. 7, can only reduce the visibility measurement error by a factor of 2.5 at best. The adaptive optics correction is lower for fainter sources and in many cases, Adaptive Optics does not significantly reduce the visibility measurement error. Therefore, the results presented in this study for $d/r_0 = 1$ are also valid for most ground-based interferometers, with or without adaptive optics. High-performance ($Strehl>0.9$) could however greatly improve the quality of the measurements.\\
\begin{figure}
\includegraphics[width=8cm]{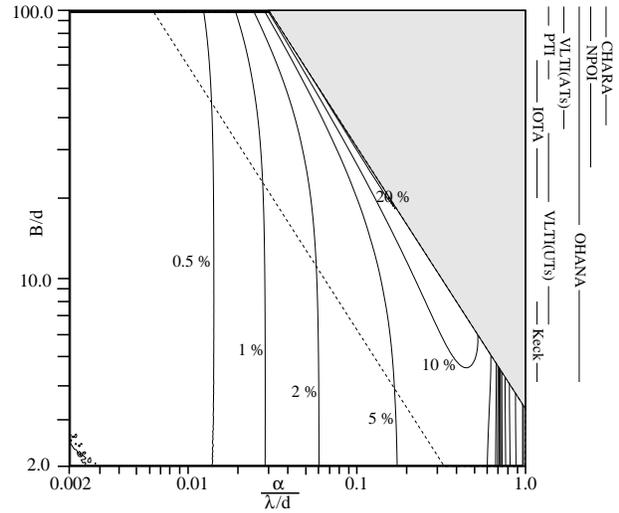}
\caption{This figure shows the relative error for measurement of the separation of the double star if the mean of the visibility measurements is used ($d/r_0=1$). The visibility of the double star is 90\% along the dashed line.}
\end{figure}
\begin{figure}
\includegraphics[width=8cm]{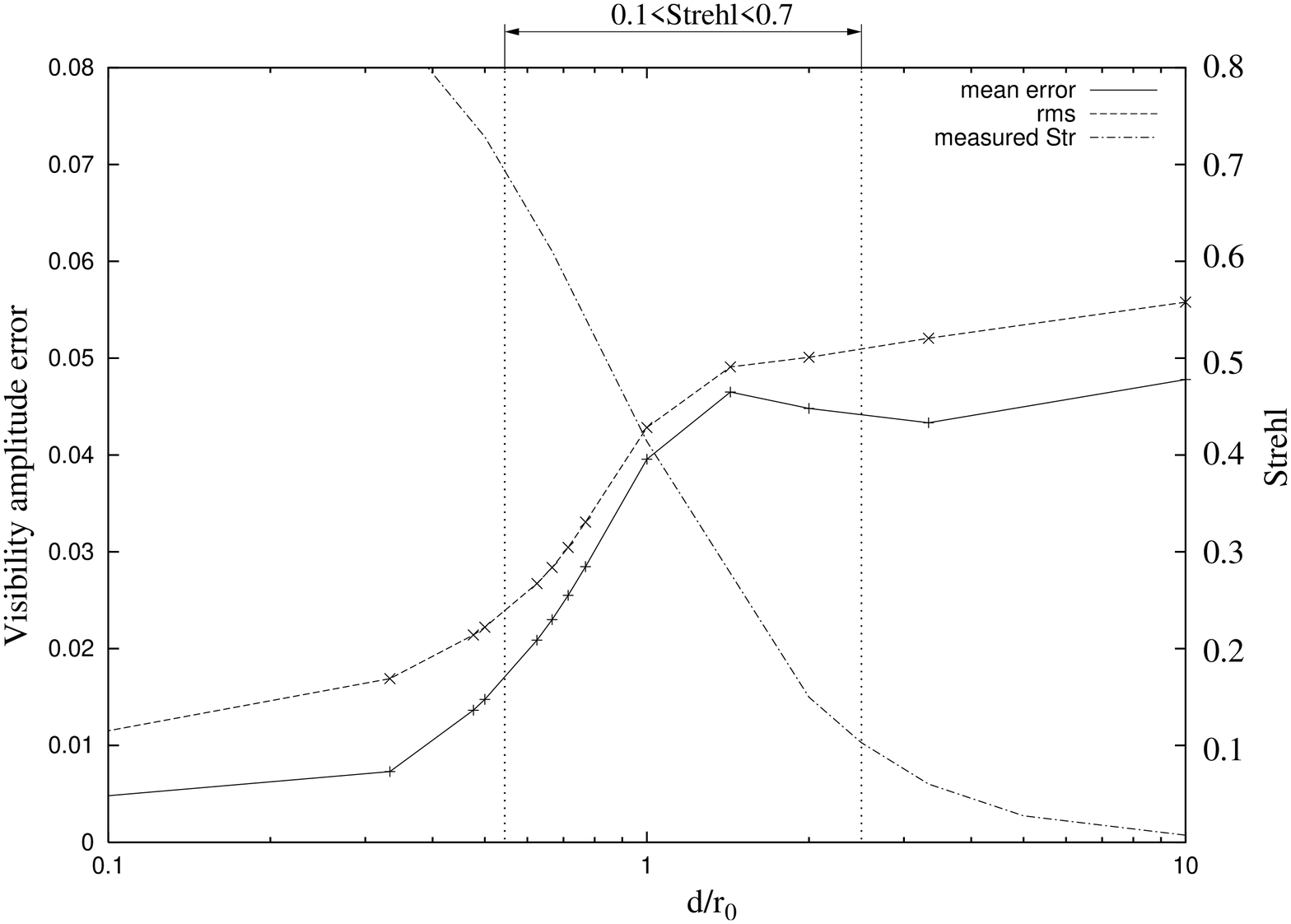}
\caption{Evolution of the systematic error in the visibility measurement and the rms of the distribution of these measurements with the strength of the atmospheric turbulence ($d/r_0$). The Strehl, as measured from the set of wavefronts used for these simulations, is also plotted. The space between the two dotted lines represents the range of residual wavefront aberration delivered by current AO systems in the near-infrared.}
\end{figure}

\subsection{Other types of sources : Stellar diameter measurement, AGNs and YSOs}
Existing interferometers can measure stellar visibilities with an accuracy of about one percent. If the star is partially resolved by a telescope of the interferometer, it is however often impossible to reach this accuracy. We consider the observation of a star of diameter $d_{star}$ with a 2-telescope interferometer. The error on the measurement of the fringe visibility due to the apparent size of the star and atmospheric turbulence have been computed with the same code as for the double star (averaging of 2000 independant photometry-corrected visibility measurements). Figure 8 shows how these errors affect the precision of the stellar diameter measurement (with $d/r_0=1$). The errors are generally smaller than in the case of the double star because the flux is more concentrated in the central region. Moreover, a very simple model (uniform disk) is adopted for the star and only one simple quantity (the diameter) is estimated. The errors tend to be large (more than a few percent) if the visibility is measured close to the first null. For $B/d > 2$, the accuracy on the stellar diameter is generally better than $1\%$ if the measure is done at baselines shorter than the resolving baseline (for which the visibility is null).\\

\begin{figure}
\includegraphics[width=8cm]{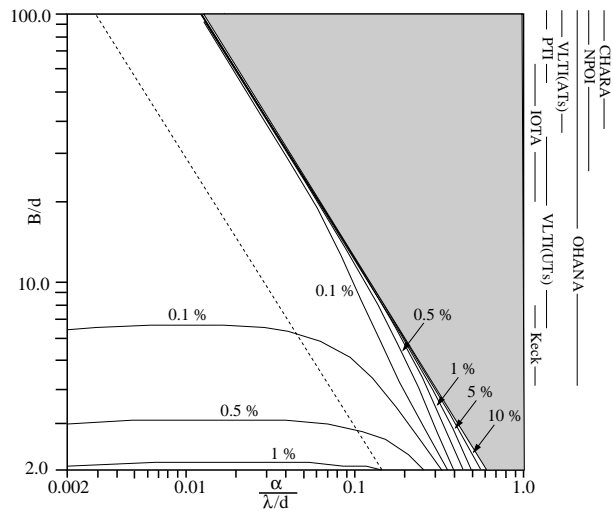}
\caption{Error in the measure of stellar diameters. $B$ is the interferometer baseline, $d$ the apertures diameter, $\alpha$ the stellar angular diameter. As for Fig. 6, the visibility along the dashed line is 90\% and $d/r_0 = 1$. The errors are minimal along a line corresponding to a 40\% visibility approximately, which explains why the 1\%, 0.5\% and 0.1\% error curves are found on either side of this line. The large errors (more than 5\%) are encountered when the fringe visibility is small.}
\end{figure}

The effect of atmospheric turbulence on the fringe visibility measurements of AGNs and YSOs is hard to quantify precisely due to the lack of a simple and accurate model of light intensity distribution for those sources. However, the two examples studied above (double stars and stellar disks) provide us with some general estimate of the measurement errors for various sources and some conclusions can be drawn:\\
\begin{itemize}
\item Atmospheric turbulence seriously alters visibility measurements of sources that are not very concentrated: stellar diameters measurements are less corrupted than double star separation measurements.
\item The error on the measured physical quantity is larger when the visibility of the object is small. Measurements near or beyond the first zero of the visibility curve should be avoided.
\item This effect still allows measurements of simple physical quantities (stellar diameters) but is likely to seriously alter more evolved measurements (small variations of stellar diameters, limb darkening, aperture synthesis imaging).
\end{itemize}

\section{Discussion}
Imaging of extended sources with fiber-fed interferometers is made very delicate because of two effects.
\begin{itemize}
\item The field of view of the single mode fiber is small ($\lambda/d$). 
\item When the source size is a significant fraction of the field of view of the single mode fiber, the visibility measurements are corrupted by atmospheric turbulence.
\end{itemize}

\paragraph{Field of view of the fiber: }

To increase the field of view of the single mode fiber, the telescope diameter could be reduced, at the cost of lowering the flux in the fibers. This solution is of limited use because it seriously affects the limiting magnitude of the interferometer. Another obvious solution is to increase the operating wavelength.\\

\paragraph{Visibility measurement errors due to atmospheric turbulence: }

This effect has been studied in \S4. One might think that because these errors can be estimated, the visibility measurements could be corrected for, using this work. Unfortunately, during actual observations, $d/r_0$ does not stay constant and additional effects (dome seeing, mirror seeing, shape of the primary mirror etc.) prevent such correction: the wavefront perturbation key parameters are not constant in time. This effect is also a function of the morphology of the source, which is generally poorly known. However, several solutions to this problem can be implemented without much complexity :
\begin{itemize}
\item {\bf Better statistical analysis of the measurements} Measurements could be ``weighted'' according to the values of the photometric outputs. For example, visibility measurements are often very corrupted when at least one of the photometric outputs is low.
\item {\bf Reducing the size of the apertures} Figures 6 and 8 show that reducing $d$ (the point of measure in these figures is then moved to the upper left, along a line of constant visibility, parallel to the dashed line) decreases the visibility measurement error. This technique is only efficient for bright (the apertures can be stopped down) sources.
\item {\bf Monitoring the PSF} Real-time imaging of the two PSFs on the fiber heads can yield to an estimate of $\rho_1(\vec{\alpha},t)$ and $\rho_2(\vec{\alpha},t)$ (equations 26 to 28).
\end{itemize}
The statistical analysis of the measurements can help with the standard deviation between measurements, but most of the systematic visibility measurement errors cannot be corrected for. Reducing the size of the aperture affects the limiting magnitude of the interferometer and degrades the precision of the measurements because of photon noise. Monitoring of the PSF seems to be the most promising technique and can be done by imaging the image reflected by the fiber head with a high-speed camera. Alternatively, efficient photon-counting detectors, such as APDs, could be used, since only a small number of ``pixels'' are required. This last technique is efficient for sources smaller than $\lambda/d$, because large sources will tend to ``blur'' the PSF (the information that leads to estimates of $\rho_1$ and $\rho_2$ is lost).\\

One elegant solution to both effects is to use several fibers per telescope. By segmenting an aperture into subapertures and using one fiber per subaperture, the field of view of the fibers is increased and the visibility measurement errors due to atmospheric turbulence are reduced. This approach offers the advantage of reducing the size of the apertures (see above) without significant loss of light.

\section{The multifiber apertures}

\subsection{The principle}
The use of one fiber per aperture limits the FOV to the size of a coherent domain in the focal plane, $\lambda/d$. Increasing the FOV can be done by increasing the size of this coherent domain (reducing $d$) or by coupling $N$ coherent domains in $N$ single-mode fibers. There are therefore two solutions to increase the FOV.\\
\begin{itemize}
\item Reducing the size of the pupil to be coupled in a fiber. To avoid losing most of the signal, the pupil is segmented into N subpupils and the light of each subpupil is coupled to a fiber.
\item Paving the focal plane with an array of fibers. Each coherent domain is coupled to a fiber. Alternatively, this technique could be implemented sequentially, by mapping the field of view with numerous pointings of the array.
\end{itemize}
Paving the focal plane with fibers would increase the FOV but poses problems for beam recombination. We have therefore chosen to use the first solution. A possible implementation of this scheme is shown on Fig. 9: a lenslet array in a pupil plane is used to couple each subaperture to its corresponding fiber. This is very similar to the optical scheme used to couple subapertures to multi-mode fibers in curvature adaptive optics systems.\\
\begin{figure}
\includegraphics[width=8cm]{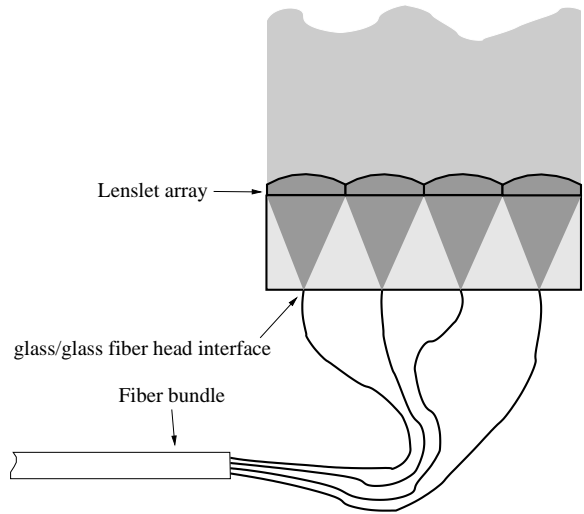}
\caption{The telescope beam is coupled in several single-mode fibers. A lenslet array in the pupil plane is used to couple each subpupil's light in a single-mode fiber.}
\end{figure}
All subapertures of the interferometer have the same physical size. As a result, the number of fibers on a given aperture is proportional to its area.\\

\subsection{Field of view}
$N$ is the number of apertures in the array, $d_{i}, i=1...N$, the diameter of each of those apertures, $d_{sa}$ the diameter of each subaperture and $M_i, i=1...N$, the number of fibers for each aperture. Since $d_{sa}$ should be the same for all subapertures, 
\begin{equation}
M_i = \left(\frac{d_i}{d_{sa}}\right)^2.
\end{equation}
The field of view of the interferometer is 
\begin{equation}
FOV = \frac{\lambda}{d_{sa}} = \sqrt{M_i} \times \frac{\lambda}{d_i}
\end{equation}
for $i = 1...N$.
At each telescope, the FOV is multiplied by the square root of the number of fibers used. Paving the focal plane with optical fibers would have resulted in the same increase of FOV.

\subsection{Coupling efficiency}
Figure 10 shows four possible subaperture geometries. We compute the coupling efficiency for each of those geometries, neglecting the edge effects ($M_i \gg 1$).\\

\begin{figure}
\includegraphics[width=8cm]{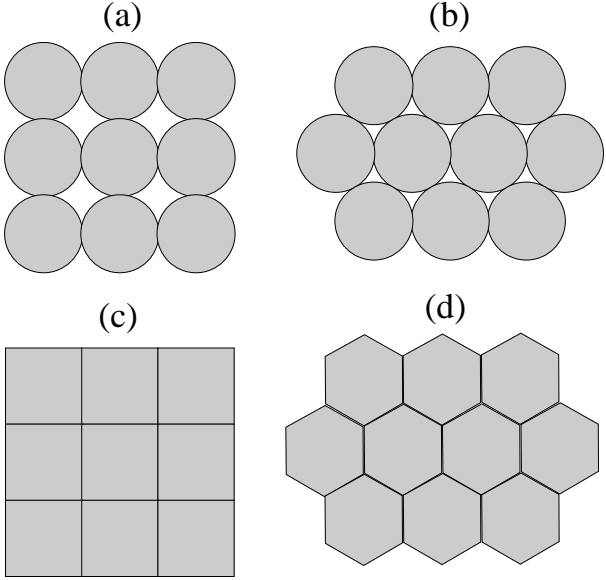}
\caption{Four possible subapertures geometries.}
\end{figure}

In geometry (a) and (b), the coupling efficiency is the product of the filling factor of the disks across the pupil and the coupling efficiency of a circular pupil into a fiber:
\begin{equation}
\rho_{a} = \frac{\pi}{4} \times 0.82 = 0.64
\end{equation}
\begin{equation}
\rho_b = \frac{\pi}{4 sin(\frac{\pi}{3})} \times 0.82 = 0.74.
\end{equation}
For geometries (c), and (d), we have computed the optimal coupling efficiency for square and hexagonal apertures using equation (1).
\begin{equation}
\rho_c =  0.793
\end{equation}
\begin{equation}
\rho_d = 0.815
\end{equation}
A tight hexagonal paving of the pupil plane allows a coupling efficiency $\rho_d$ almost (0.5 \% difference) as good as a direct coupling of the entire circular pupil into a single fiber. For apertures with a central obstruction of diameter $a \times d$, equation (2) becomes
\begin{eqnarray}
\lefteqn{\rho(\alpha) = \frac{8}{\omega^2} e^{-2\left(\frac{\alpha f}{\omega}\right)^2} \times }\nonumber \\
& \hspace*{-0.5cm} \left[ \int e^{-\left(\frac{r}{\omega}\right)^2} I_0(2r\alpha f/\omega^2) \left[J_1(\pi dr/\lambda f) - a J_1(a \pi dr/\lambda f)\right] dr \right]^2.\hspace*{0.2cm}
\end{eqnarray}
\begin{figure}
\includegraphics[width=8cm]{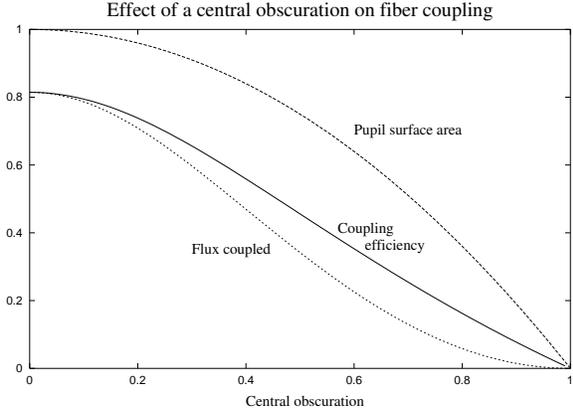}
\caption{The central obscuration of a pupil reduces the coupling efficiency. The pupil surface area is normalized to the unobstructed pupil surface area. The flux coupled is normalized to the total flux gathered by the unobstructed aperture.}
\end{figure}
Using this equation, we can compute, for each value of $a$, the combination of $\omega$, $f$ and $d$ which yields the best coupling efficiency. As can be seen on Fig. 11, the coupling efficiency decreases rapidly with the size of the central obscuration. For a central obstruction of 0.4, the coupling efficiency is 56\%. However, if the pupil is subdivided into hexagonals subapertures, each coupled to a fiber, the overall coupling efficiency is $\rho_4 = 0.815$.\\
Under diffraction-limited conditions, the division of the pupil in several hexagonal subpupils does not result in a loss of coupling efficiency. In the case of a circular pupil with a central obstruction, there is a gain of coupling efficiency.\\

\subsection{Beam combining}
The high number of optical fibers in this concept is a strong constraint for the design of a beam combiner. Two possible schemes have been identified.
\begin{itemize}
\item {\bf Reconstruction of the entrance pupils.}
The most intuitive approach is to optically reconstruct the incoming wavefront (in fact, some approximation of it) for each pupil of the interferometer, as seen on Fig. 12. Wide-field beam recombination techniques can then be used as if each entire telescope beam had been transported. In this scheme, the fibers (several per telescope) are only used to transport the wavefront from the telescopes to the beam combiner. One attractive approach for beam combination is a Fizeau-type beam combiner (shown on Fig. 12), where the beams are combined in a common focal plane. The pupil of the beam combiner can be identical to the entrance pupil, it can be a densified (\cite{lab96}) version of it, or it can be a fixed pupil. In the first 2 cases, the beam combiner pupil needs to be rearranged to follow the geometry of the entrance pupil of the array (as seen by the source). 
\begin{figure}
\includegraphics[width=8cm]{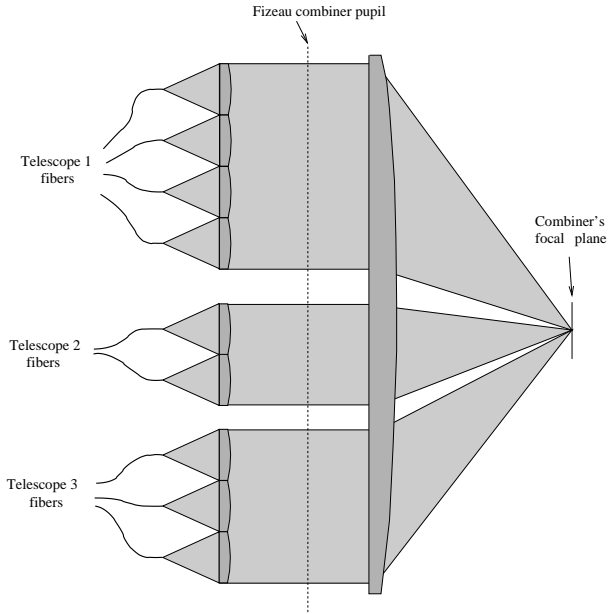}
\caption{One possible scheme for beam combination: the telescopes' pupils are optically reconstructed and fed in a Fizeau combiner.}
\end{figure}
\item {\bf Use of integrated optics components.} 
A more promising technique is the use of a large number of fiber couplers to determine the complex amplitude of each subpupil in the array, relative to a common (for the whole array) phase. Phase closure techniques could be used between subapertures of different telescopes if more than 2 telescopes are used. Integrated optics components (\cite{mal99,ber99,ber01}) can include a high number of fiber couplers in a small volume and seem to be a very promising technology for this beam combination scheme, illustrated in Fig. 13. In this case, the beam combiner considers the fibers as the coherent beams from a high number of sub-telescopes (as many as there are fibers). The beams of sub-telescopes belonging to the same telescope are already cophased thanks to adaptive optics. It would be very interesting to determine what is the optimal set of beam couplings for wide-field imaging with a limited number of photons.  
\begin{figure}
\includegraphics[width=8cm]{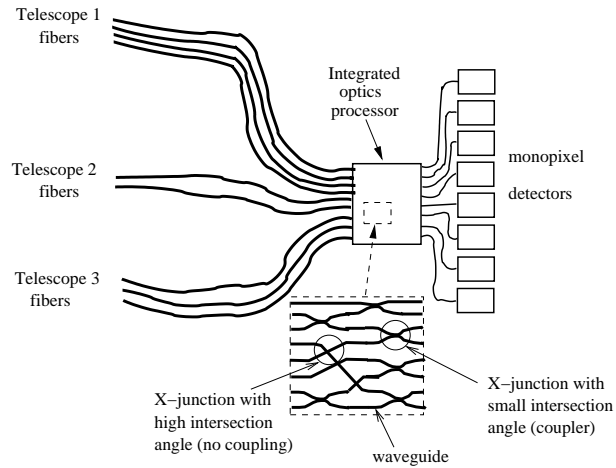}
\caption{Another possible scheme for beam combination: multiple interferences are done between the fibers inside an integrated optics ``optical processor''.}
\end{figure}
\end{itemize}

\subsection{Photometric calibration}
In this concept, the coupling field of view of an individual fiber is increased by dividing the entrance pupil into subapertures, each of those being coupled to a fiber. When this coupling field of view is significantly larger than the observed source, the effect presented in \S3 and \S4 is greatly reduced. If this coupling field of view is significantly larger than the atmospheric seeing, the flux in each fiber is constant in time and there is no need for photometric calibration of the interferometer.

In practice, this second requirement is hard to meet because of the very large number of fibers required. Therefore, photometric calibration of the interferometer would still greatly improve the fringe visibility measurement accuracy, as it currently does for monofiber interferometers (FLUOR instrument on the IOTA interferometer). This photometric calibration of the interferometer would be done by spitting the light of each fiber before the beam combination, yielding one photometric output per subaperture. Such a calibration unit could be integrated directly into the integrated optics processor presented in \S6.4. It is very interesting to note that in this scheme, the different subapertures can be considered as different telescopes of an interferometric array, regardless of their belonging to the same main aperture.

\section{Use on a space interferometer}

\subsection{(u,v) plane coverage and field of view.}
Guyon \& Roddier (2001) have demonstrated that with only six 2-meter apertures, a full (u,v) plane coverage up to a 60m baseline is accessible with the rotation of an array. Wide field of view imaging is possible with such arrays if the phase information across the pupil is preserved: collapsing each pupil on a single mode fiber would reduce the field of view. To exploit this full (u,v) plane coverage and reconstruct wide field images, the flux received by the telescope over a large solid angle needs to be brought to the beam combiner. We consider such a optimized array geometry and review the characteristic field of view quantities for a fiber interferometer using rotational aperture synthesis. In this section, we consider a array of $N$ identical telescopes: $d_0=d_1=...=d_N=d$ and $M_0=M_1=...=M_N=M$.

\begin{itemize}
\item{\bf Coupling field of view}\\
Using the results of \S6.2, 
\begin{equation}
FOV_c =  \frac{\lambda}{d_{sa}} = \sqrt{M} \times \frac{\lambda}{d},
\end{equation}
with  $d$, the diameter of each telescope, $d_{sa}$ the diameter of each subaperture and $M$, the number of fibers for each aperture.

\item{\bf Fourier field of view}\\
We first consider an array of N telescopes of infinitely small diameter. There are $N(N-1)+1$ (u,v) points measured per snapshot exposure (two symmetrical points per baseline plus the origin). In the case of a rotating array, the (u,v)-plane coverage of the final reconstructed image is a series of $N(N-1)/2$ circles (excluding the origin). This correctly describes the (u,v) plane coverage if each aperture's flux is coupled into a fiber (the information about the size of the pupil is not preserved). The maximum distance between two consecutive baseline lengths measured is
\begin{equation}
d_c = e(N) \frac{B}{N(N-1)/2}
\end{equation}
where $e(N)$ is the efficiency of the (u,v) plane coverage of the array (\cite{guy01}) and $B$ is the longest baseline of the array. This efficiency decreases from 1 to 0.869 as $N$ increases from 4 to 10. In this simple estimate, we will consider $e(N) = 1$, which is a valid approximation for $N<10$. In a (u,v)-plane coverage obtimized array, the diameter or the individual apertures is sufficient to fill the ``gap'' between consecutive baselines: $d=d_c$ (\cite{guy01}).

Reconstructing an image requires one to interpolate the (u,v) plane between the measured spatial frequencies: this is equivalent to a convolution in the (u,v) plane by a kernel of characteristic size equal to the distance between the known (u,v) points. In this case, this distance is $d_c$. Therefore, with only one fiber per aperture, this imposes a limitation $FOV_f$ on the field of view of the reconstructed image:
\begin{equation}
FOV_f = \frac{\lambda}{d_c} = \frac{\lambda N(N-1)}{2B}.
\end{equation}
If no interpolation is done in the (u,v) plane, there is no ``windowing'' in the image plane, but the PSF has strong secondary peaks separated by $FOV_f$: the clean field of view would therefore be $FOV_f$. In the case of a subdivision of each pupil into $M$ hexagonal subpupils ($M$ fibers per aperture), equation (37) becomes
\begin{equation}
d_c < \frac{e(N) \frac{B}{N(N-1)/2}}{\sqrt{M}}
\end{equation}
and equation (38) becomes
\begin{equation}
FOV_f > \sqrt{M} \frac{\lambda N(N-1)}{2B}.
\end{equation}
\end{itemize}
From equations (36) and (40), and noting that, for an array with a full (u,v) plane coverage ($d=d_c$), with $e(N) \approx 1$, equation (37) yields $N^2d \approx  2B$,
\begin{equation}
FOV_f > FOV_c.
\end{equation}
Therefore, on the reconstructed image, the field of view will be limited by the coupling of the light into the fibers, and
\begin{equation}
FOV = \sqrt{M} \times \frac{\lambda}{d}.
\end{equation}
Equation (42) states that the number of (u,v) points measured is sufficient to yield a clean PSF without strong secondary peaks inside the field of view permitted by the fiber coupling.

\subsection{Spatial filtering by the fibers}
Single-mode fibers have been proposed for use on space interferometers because of their spatial filtering capabilities. The wavefront errors at the telescope's pupil result in a loss of coupling efficiency into the fiber but there is no loss of coherence inside the fiber: it only allows one coherent mode to be coupled. Because image reconstruction and nulling are more tolerant to amplitude variations than phase variation, the constraints on the wavefront quality are reduced by the use of such fibers.

In this concept, the wavefront over each subaperture is cleaned by this spatial filtering and the coupling efficiency decreases with the amount of wavefront distortion over this subaperture. The spatial filtering is done at a higher spatial frequency than if only one fiber was used per aperture. The signals in the fibers are perfectly cophased only if the wavefront is flat at spatial frequencies lower than $1/d_{sa}$: it is therefore essential to accurately control the quality of the wavefront at low spatial frequencies, either before injection into the fibers (adaptive optics) or after (active delay lines to cophase the fibers). Spatial filtering transforms the wavefront distortions of higher spatial frequencies into coupling efficiency losses, which can also affect the the visibility measurements. This effect can be corrected for by photometric calibration of the interferometer, as discussed in \S6.5.

\section{Use on a ground-based interferometer : OHANA}
\subsection{Presentation of the OHANA project}
OHANA (Optical Hawaiian Array for Nanoradian Astronomy) makes use of the existing optical and near-infrared telescopes on the summit of Mauna Kea to build a fiber interferometer (\cite{mar96,mar98}). Optical fibers are preferred to classical beam propagation for the simplicity of the beam collection at the telescope foci and the easy transport of coherent light over hundreds of meters. The spatial filtering is also very desirable since the Strehl ratios delivered by the adaptive optics systems on the OHANA telescopes, in the near-infrared, are typically about 50\% or less.
\begin{table}[h]
{\small
\begin{tabular}{|c|c|c|c|c|}
\hline
Telescope & D(m) & Lat.(d) & Long.(d) & elev(m)\\ 
\hline
CFHT & 3.6 & 19.8252518 & 155.468876 & 4204.1\\
Gemini & 8.1 & 19.8238014 & 155.469047 & 4213.4\\
Keck 1 & 10.0 & 19.8259465 & 155.474719 & 4159.6\\
Keck 2 & 10.0 & 19.8265606 & 155.474234 & 4159.6\\
Subaru & 8.2 & 19.8255040 & 155.476019 & 4163.0\\
IRTF & 3.0 & 19.8262183 & 155.471999 & 4168.1\\
UKIRT & 3.8 & 19.8224315 & 155.470327 & 4198.5\\
UH & 2.2 & 19.8229911 & 155.469434 & 4213.6\\
\hline
\end{tabular}
}
\caption{Telescopes included in the simulation of the OHANA array, and their 3D coordinates.}
\end{table}

\subsection{Fiber coupling Field of View}
The OHANA array, because it makes use of pre-existing telescopes, is made of apertures of various sizes, from the 10m-class telescopes (Keck I, Keck II, Subaru and Gemini) to the 4m-class telescopes (CFHT, IRTF and UKIRT).  The field of view of the interferometer is the smallest FOV seen by the fibers. If only one fiber per telescope is implemented, the field of view is limited to 20 milli-arc-second at $\lambda = 1 \mu m$ and 40 milli-arc-second at $\lambda = 2 \mu m$.\\
A more serious problem for the imaging of extended objects is the matching of the field of view for the different telescopes studied in \S3.3: with only one fiber per telescope, the field of view coupled into the fibers varies by a factor of 2 from telescope to telescope (factor of 4 in area). This is a serious problem when the observed source overfills the smallest field of view. Another obstacle to high accuracy object visibility measurements is the effect of residual wavefront aberrations, studied in \S4.

The use of several fibers per aperture, as presented in \S6, would solve these two problems and allow observations of extended objects. Table 2 lists, for each telescope of the OHANA array, the approximate number of fibers required for an observation with a specific FOV.\\ 
\begin{table}[h]
{\small
\begin{tabular}{|c|c|c|c|c|}
\hline
 & 0.05'' & 0.1'' & 0.5'' & 1''\\ 
\hline
CFHT & 1 & 4 & 77 & 305\\
Gemini & 4 & 16 & 377 & 1505\\
Keck 1 & 6 & 24 & 588 & 2351\\
Keck 2 & 6 & 24 & 588 & 2351\\
Subaru & 4 & 16 & 396 & 1581\\
IRTF & 1 & 3 & 53 & 212\\
UKIRT & 1 & 4 & 85 & 340\\
UH & 1 & 2 & 29 & 114\\
\hline
\end{tabular}
}
\caption{Approximate number of fibers required per telescope of the OHANA array for 0.05, 0.1, 0.5 and 1 arcsecond FOV observations at 1 $\mu$m.
}
\end{table}

\subsection{(u,v) plane coverage}
The instantaneous (u,v) plane coverage of the OHANA array is very sparse due to the large baselines and relatively small number of apertures. However, aperture supersynthesis greatly improves this situation: the measured (u,v) points lie along arcs that are drawn as the object moves across the sky.\\ 
\begin{figure*}
\centering
\includegraphics[width=15cm]{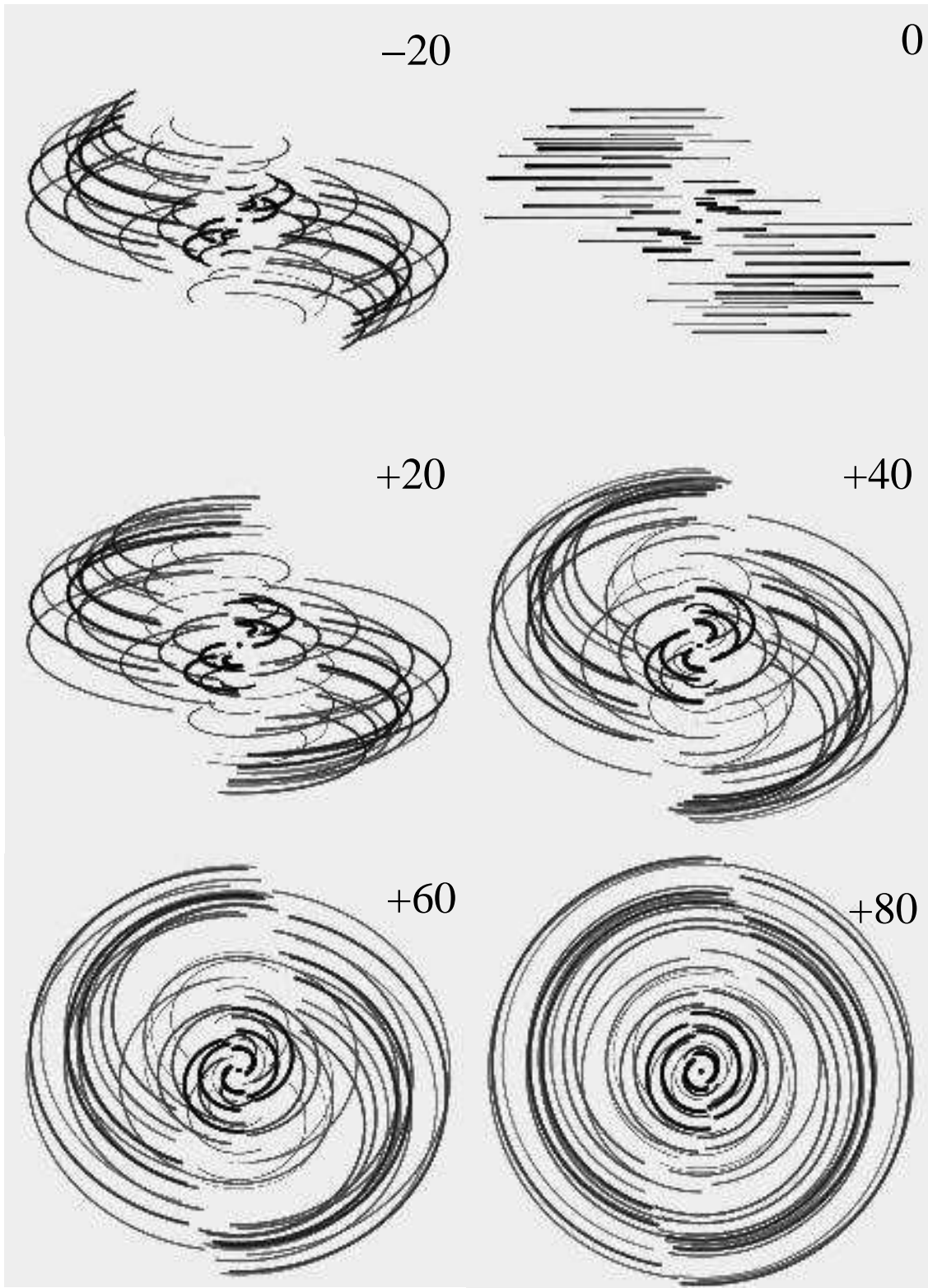}
\caption{OHANA maximum (u,v) coverage for objects at various declinations.}
\end{figure*}

The size of the telescopes of the array (up to 10 meters) is not negligible when compared to the distance between them (up to 800m). Therefore, a significant gain in (u,v) plane coverage can be achieved by preserving the phase information across the pupil. The points measured in the (u,v) plane are then thick arcs. Figure 14 shows the (u,v) coverage accessible to OHANA when observing sources at various declinations, when their elevation is more than 20 degrees above the horizon. Although there is a significant gain in (u,v) plane coverage, the number of telescopes and their diameters are still too small to fill most of the frequency gaps between the arcs. The Fourier clean field of view, which is given by the spacing between the arcs, is about $\lambda/20m$ for OHANA, which is about twice as small as the field of view of the fibers of the big telescopes of the array. Therefore, in this case, the use of several fibers per aperture is mostly motivated by the need to reduce the turbulence-induced visibility measurement errors in the array and increase the coupling FOV: the size of the gaps in the (u,v) plane coverage is not significantly reduced by using several fibers per telescope.

\section{Conclusion}
Wide field imaging with fiber-fed interferometers (and more generally with interferometers which use spatial filtering) is made very difficult because of the small field of view of the fibers (or the pinhole). This work has also demonstrated that the spatial filtering of partially corrected turbulent wavefronts limits the fringe visibility measurement accuracy.
 This effect is especially serious for the next generation of interferometers (Keck, VLTI, OHANA) which will make use of large (10m-class) telescopes with adaptive optics. For such projects, this work has shown that photometry-calibrated fringe visibility measurement errors will commonly be of the order 5\% or more for extended objects (AGNs and YSOs) if only one fiber per telescope is used.
The use of several fibers per aperture is an attractive solution to both improve the photometry-corrected visibility measurement accuracy and extend the field of view of these interferometers, and it can also allow fiber-based wide field interferometric imagers to operate from space.

\begin{acknowledgements}
The author is very grateful to Olivier Lai, Guy Perrin and Katherine Roth for constructive discussions about this work and suggestions to improve the manuscript. 
\end{acknowledgements}

\end{document}